\documentclass[aps,prl,twocolumn,superscriptaddress,showpacs]{revtex4}
\usepackage{graphicx}
\usepackage{mathrsfs}
\usepackage{bm}

\begin{document}

\title{Magnetic and Electronic Properties of Metal-Atom Adsorbed Graphene}

\author{Jun Ding$^\dag$}
\affiliation{Institute of Physics and Beijing National Laboratory
for Condensed Matter Physics, Chinese Academy of Sciences, Beijing
100190, China}
\author{Zhenhua Qiao$^\dag$}
\affiliation{Department of Physics, The University of Texas at
Austin, Austin, Texas 78712, USA}
\author{Wanxiang Feng}
\affiliation{Institute of Physics and Beijing National Laboratory
for Condensed Matter Physics, Chinese Academy of Sciences, Beijing
100190, China}
\author{Yugui Yao$^\ddag$}
\affiliation{Institute of Physics and Beijing National Laboratory
for Condensed Matter Physics, Chinese Academy of Sciences, Beijing
100190, China}
\author{Qian Niu$^\S$} \affiliation{Department of
Physics, The University of Texas at Austin, Austin, Texas 78712,
USA}\affiliation{International Center for Quantum Materials, Peking
University, Beijing 100871, China}
\begin{abstract}
We systematically investigate the magnetic and electronic properties of graphene adsorbed with diluted 3$d$-transition and noble metal atoms using first principles calculation methods. We find that most transition metal atoms (i.e. Sc, Ti, V, Mn, Fe) favor the hollow adsorption site, and the interaction between magnetic adatoms and $\pi$-orbital of graphene
induces sizable exchange field and Rashba spin-orbit coupling, which together open a nontrivial bulk gap near the Dirac points leading to the quantum-anomalous Hall effect. We also find that the noble metal atoms (i.e. Cu, Ag, Au) prefer the top adsorption site, and the dominant inequality of the AB sublattice potential opens another kind of nontrivial bulk gap exhibiting the quantum-valley Hall effect.
\end{abstract}
\date{\today{}}

\maketitle

\emph{Introduction ---.} After an initial rush~\cite{RMPgraphene}
since its first experimental exfoliation~\cite{Geim2004}, graphene research has turned
toward the application of graphene-based
electronics~\cite{RMPgraphene,GraApplication}. Graphene itself is a
zero-gap semiconductor characterized by the linear Dirac-type
dispersion, which is closely related to many amazing properties,
e.g. high electron mobility and half-integer quantum Hall
effect (per spin and valley). However, a band gap in graphene is highly desirable for
designing semiconductor devices. So far, there are several proposals
of engineering a band gap, such as by employing a staggered AB
sublattice potential~\cite{DiXiao}, strain effect~\cite{strain}, intrinsic spin-orbit coupling~\cite{Kane} in single layer graphene,
or applying a perpendicular electric field in bilayer graphene~\cite{BilayerGap}.

Graphene is also considered as a promising candidate for
spintronics, which generally require imbalanced spin up and down carrier populations. However, the
pristine graphene is nonmagnetic, and one need to employ
external methods to magnetize it, e.g. by decorating hydrogen,
sulfur or metal adatoms on
graphene~\cite{MagnetizedHSF,Cohen,Savini}, or substituting metal
impurities for carbon atoms in graphene~\cite{substitution}. Alternatively,
spin-orbit coupling is thought to be another source for manipulating the spin degrees of
freedom through electric means~\cite{SO-spintronics}. In graphene there are two kinds of spin-orbit couplings: intrinsic and
extrinsic~\cite{Kane}. The former one has been shown to be unrealistically
weak~\cite{WeakSOI}. The latter one, which is also known as Rashba spin-orbit
coupling, arises from the top-bottom layer symmetry breaking due to
the presence of a substrate or strong perpendicular electric field, which was reported to be remarkably large in graphene placed on top of
Ni(111) surface~\cite{Rashba}.
\begin{figure}
\includegraphics[width=6.6cm,totalheight=3.5cm,angle=0]{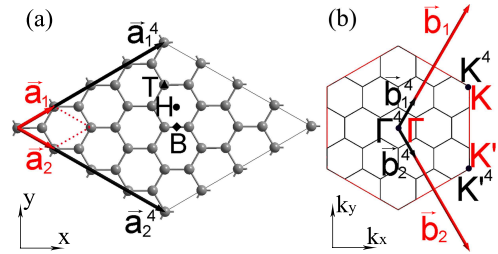}
\caption{(Color online) (a) A $4\times 4$ supercell of graphene.
$\vec{a}_{1,2}$ ($\vec{a}^4_{1,2}$) indicates the primitive vectors
for the $1\times 1$ ($4\times 4$) supercell. Three possible
adsorption sites of single adatom on graphene are labeled as:
hollow~(H), top~(T), and bridge~(B). (b) Corresponding reciprocal
momentum space structures: red and black vectors are for the
$1\times 1$ and $4\times 4$ supercells, respectively.} \label{Fig1}
\end{figure}

\begin{table*}
\begin{ruledtabular}
\caption{Energetic and structural properties of 12 metal atoms
located at the lowest energy adsorption sites on the $3\times3$ and
$4\times4$ supercells of graphene. The properties listed include the
favored adsorption sites, adatom-graphene distance $d$~(${\rm
\AA}$), adsorption energy $\delta E$~(in units of $eV$/atom), bulk
gap $\Delta$~($meV$), charge transfer $\delta Q$~($e$) from adatoms
to graphene, and the magnetization of the adatoms
$\mu_{ag}$~($\mu_B$) on the $4\times4$ supercell. Note that data
inside the parentheses are used for comparison at the hollow site.
$0.0^*$ means that the global band gap is zero, but the local gap is
finite around $K/K'$ points for the $4\times4$ supercell or $\Gamma$
points for the $3\times3$ supercell.}
\begin{tabular}{cccccccccccc}
& \multicolumn{5}{c}{$3\times3$ supercell}  & \multicolumn{6}{c}{$4\times4$ supercell}   \\
\cline{2-6} \cline{7-12}
Atoms & site & $d$ & $\delta$E &$\Delta$ &$\delta Q$ & site  &  $d$ & $\delta$E &$\Delta$ & $\delta Q$ &$\mu_{ag}$  \\
\hline
Sc    &H&  1.92 &1.26 & 2.5 & 0.97 & H& 1.97 & 1.43 & 2.5 & 0.89 & 2.30 \\
Ti    &H&  1.86 &1.76 & 0.0$^*$ & 0.88 & H& 1.86 & 1.75 & 1.4 & 0.87 & 3.41   \\
V     &H&  1.89 &0.94 & 0.0$^*$ & 0.67 & H& 1.87 & 1.01 & 2.2 & 0.71 &  4.51    \\
Cr    &B(H)&  2.33(2.17) &0.14(0.13)& 0.0($0.0^*$) & 0.35(0.41)& B(H)&  2.37(2.15) &0.16(0.15)& 0.0$^*$(2.5) & 0.37(0.45) & 5.64(5.52)     \\
Mn    &H&  2.07 &0.14 & 4.5 & 0.56 & H& 2.09 & 0.14 & 4.5 & 0.57 & 5.42     \\
Fe    &H&  1.56 &0.98 & $0.0^*$ & 0.81 & H& 1.56 & 0.95 & 5.5 & 0.81 & 1.97   \\
Co    &H&  1.55 &1.31 & 0.0$^*$  & 0.63   & H& 1.53 &1.67 & 0.0$^*$ & 0.66 & 0.99  \\
Ni    &H&  1.58 &1.37 & 60 & 0.51 &  H &  1.58  &1.39 & 0.0  & 0.52    &  0.00   \\
Cu    &T(H)&  2.24(2.17) &0.24(0.10) & 0.7(0.4)   & 0.16(0.14)    & T(H)&  2.25(2.17)  &0.25(0.10) &  38.8(7.0)   & 0.15(0.15)    &  0.90(0.92)    \\
Zn    &H&  3.78 &0.03 &0.6 &0.02 & H&  3.78  &0.02 &0.0 &0.02 & 0.00 \\
(LDA) &H&  3.04 &0.14  &13.9 &0.03&H &  3.02  &0.14 &0.0 &0.03& 0.00 \\
Au    &T(H)&  3.13(3.47) & 0.14(0.13)  &0.0$^*$(2.9) & -0.08(-0.09) & T(H)&  3.45(3.54)  &  0.16(0.16) & 0.6(1.7) &-0.02(-0.15) & 0.84(0.84) \\
(LDA) &T(H)&  2.37(2.43) &1.20(0.94) &  9.4(2.0) &-0.06(-0.05) &T(H)  &  2.40(2.52)  &0.77(0.50)& 161.4(8.6) &-0.09(-0.09) & 1.03(0.96) \\
Ag    &T(H)&  3.57(3.69) & 0.03(0.03)  & 0.0(1.5) & 0.04(0.02) & T(H)&  3.49(3.58)  &    0.03(0.03)& 3.0(2.3) &0.02(0.02)& 1.00(1.00)\\
(LDA) &T(H)&  2.45(2.52) &0.82(0.76)& 2.8(6.8) &0.08(0.07) & T(H) &  2.44(2.51)  &0.38(0.31)  & 21.8(8.7) &0.11(0.08)& 0.87(0.93)\\
\end{tabular}\label{table1}
\end{ruledtabular}
\end{table*}

In this Letter, we present a systematic investigation on the
magnetic and electronic properties of graphene adsorbed with
3$d$-transition and noble metal atoms including the spin-orbit
coupling using first principles calculation methods. The results show that most transition metal adatoms favor the
\emph{hollow} site, while the noble metal adatoms prefer the
\emph{top} site. We find that both types of adsorption give rise
to nontrivial bulk gaps. Further Berry-phase
analysis~\cite{NiuRMP} demonstrates that the gap opening mechanisms
are completely distinct, i.e. one arises from the joint
effect of exchange field and Rashba spin-orbit coupling leading to
the quantum-anomalous Hall~(QAH) effect, while the other one
originates from the inequality of the AB sublattice potential resulting in
the quantum-valley Hall~(QVH) effect. Our findings should not only
provide a new scheme for the gap opening in graphene, but also show
great potential for the realization of the long-sought quantum
anomalous Hall effect and the dissipationless valleytronics.

\emph{Computational Methods ---.} The adatom-graphene system is
modeled using one metal atom on top of a $N \times N$ ($N$=3,~4)
supercell of graphene. As illustrated in Fig.~\ref{Fig1}(a), a supercell
is composed of $2N^2$ carbon atoms and one metal atom, and the
single atom has three possible adsorption sites: hollow~(H),
top~(T), and bridge~(B). Panel (b) plots the Brillouin-zone of
the $4 \times 4$ supercell comparing to that of the $1 \times 1$
supercell with well-defined valley indices $K$ and $K'$.

The first principles calculations were performed using the
projected-augmented-wave method~\cite{PAW} as implemented in the
Vienna Ab-initio Simulation Package~(VASP)~\cite{VASP}. The
generalized gradient approximation (GGA) exchange-correlation
functional~\cite{GGA} was mainly used except those specified by the
local density approximation (LDA) in Table~\ref{table1}. The kinetic
energy cutoff was set to be 500 eV, and the experimental lattice
constant of graphene $a=2.46~{\rm \AA}$ was used. During the
structure relaxation, all atoms were allowed to relax along the
normal direction of graphene and all parameters were chosen to
converge the forces to less than 0.01~$\rm eV/{ \AA}$. The first
Brillouin-zone integration was carried out by using the
$12\times12\times1$ and $6 \times 6 \times1$ Monkhorst-Pack grids
for the $3 \times 3$ and $4 \times 4$ supercells, respectively. A
vacuum buffer space of 15~{\rm \AA} was set to prevent the
interaction between adjacent slabs.

\emph{Adsorption Analysis ---.} Table \ref{table1} summarizes the
stable adsorption sites, energetic and structural properties of 12
kind of 3$d$-transition and noble metal atoms for the $3\times3$ and
$4\times4$ supercells. The stable adsorption site, distance $d$ and
magnetization of adatoms $\mu_{ag}$ were obtained by only considering
the magnetization, while other quantities were calculated by further
including the spin-orbit coupling. The adsorption energy is defined
as: $\delta E=E_{a}+E_{g}-E_{ag}$, where $E_{a}$, $E_{g}$ and
$E_{ag}$ are the energies of the isolated atom, $N\times N $
supercell of graphene, and adatom-$N\times N$ supercell of graphene,
respectively. The charge transfer $\delta Q$ was calculated based on
the Bader charge analysis~\cite{chargeTransfer}. For clear
demonstration, we divide the adsorption analysis into two parts:

(a) For the 3$d$-transition metal adatoms, we observe that most of
them except Cr favor the hollow site.
(i) For Sc, Ti, V, Fe, Co, and Ni, the
adsorption distances are very short (less than $2~\rm {\AA}$), and
the adsorption energy $\delta E$ and the charge transfer are very
large, which together indicate a strong hybridization; (ii) For Cr
and Mn, the adsorption distance is slightly larger than $2~\rm
{\AA}$, the adsorption energy is relatively weak but the charge
transfer is still remarkably large, which corresponds to a moderately weak
adsorption; (iii) For Zn, both the large adsorption distance and
extremely small adsorption energy and charge transfer signal an extremely weak
hybridization. In the following, we show that the reconstruction of the electronic configurations of the adsorbed metal
atoms is intimately dependent on the strength of adsorption on the graphene sheet.

In Table~\ref{table1}, we notice that the magnetic moments of
adatoms $\mu_{ag}$ are greatly altered compared to that of the
isolated atoms. This is found to obey the following rules: (1) When the 3$d$-shell is less than half filled~(i.e.
Sc, Ti, V), the strong hybridization lowers the energy of the
3$d$-orbital. This makes the 4$s$ electrons transfer to the
unoccupied 3$d$-orbital. Since the resulting 3$d$-shell is not over
half-filled, the Hund's rule dictates that all the 3$d$ electrons possess
the same spin-polarization leading to the increase of the magnetic
moments $\mu_{ag}$; (2) When the 3$d$-shell is exactly
half-filled~(i.e. Cr, Mn), the adsorption is weak therefore
the 3$d$-orbital energy is less affected during the hybridization.
Based on the Hund's rule, all 3$d$-electrons are equally
spin-polarized showing a maximum magnetization. In the influence of
the high magnetization from the 3$d$-shell, the 4$s$ electrons tend
to align with the spin of the 3$d$ electrons, which slightly increases
the magnetic moment; (3) When the 3$d$-shell is over
half-filled~(i.e. Fe, Co, Ni), the strong hybridization
lowers the 3$d$-shell energy and the transfer of 4$s$ electrons to
3$d$-orbital decreases the magnetic moment due to the occupying of
the unpaired 3$d$-orbital~\cite{Cohen}. In particular, for Ni the
two 4$s$ electrons are transferred to $3d$-orbital and form a closed
3$d$ shell, giving rise to a vanishing magnetic moment; (4) For
atoms with closed 3$d$ and 4$s$ shells (i.e. Zn), it behaves
like a inert atom leaving graphene nearly unaffected.

(b) For the noble metal adatoms, we find that they are stable at the
top adsorption site. Consistent with Ref.~[\onlinecite{Cohen}], the long adsorption distance, low adsorption energy, and weak charge transfer originate from the physical adsorption involving the van de Waals forces. In Table~\ref{table1}, we also provide the LDA results for Au and Ag adsorption. The major differences are that the adsorption distance $d$ is dramatically decreased and the adsorption energy $\delta E$ is exponentially increased. Because of the closed $d$ shells, the single $s$ valence electron contributes to the magnetic moment $\mu_{ag}\sim1\mu_B$.

Due to the magnetic proximity effect with the magnetic adatoms, graphene can be magnetized. In the
absence of spin-orbit coupling, we estimated the induced exchange field~\cite{exchangefield} near the Dirac
points to be $\lambda$=145~(Sc), 277~(Ti), 199~(V),
359~(Cr), 181~(Mn), 57~(Fe), 462~(Co), 0~(Ni), 51~(Cu) in units of
meV for the hollow-site adsorption on the $4\times4$ supercell.
\begin{figure}
\includegraphics[width=8.5cm,totalheight=13cm,angle=0]{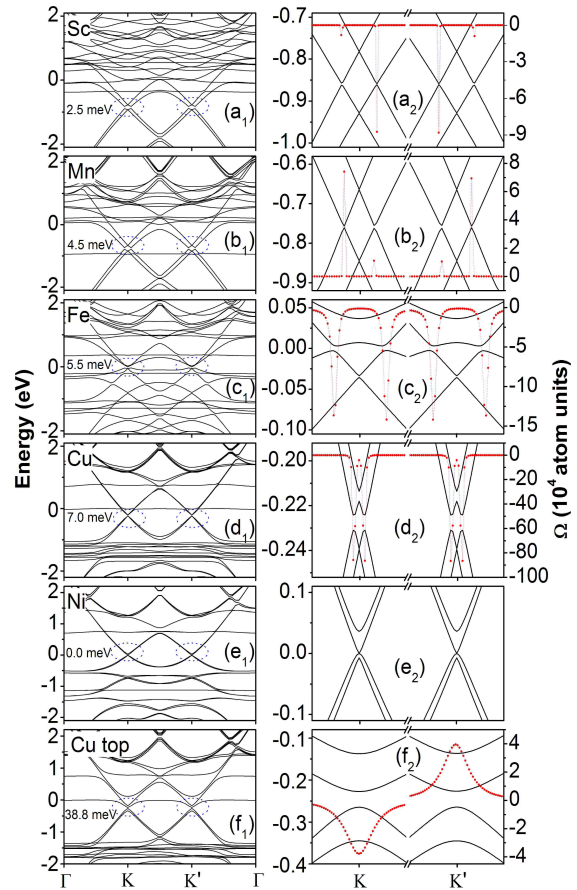}
\caption{(Color online) Panels ($\rm a_1$)-($\rm e_1$): Full bulk band structures
based on GGA for Sc ($\rm a_1$), Mn ($\rm b_1$), Fe ($\rm c_1$), Cu ($\rm d_1$), Ni ($\rm e_1$) adsorbed at
the hollow site on a $4 \times 4$ supercell of graphene along high
symmetry lines by including the spin-orbit coupling. Panel ($\rm f_1$): Full
bulk band structure based on GGA for Cu adsorbed at top site on a $4
\times 4$ supercell of graphene. Panels ($\rm a_2$)-($\rm f_2$): Zooming in of the
bands around $K$ and $K^\prime$ Dirac points (in black), and the
Berry curvatures $\Omega({\bm k})$ for the whole valance bands (in
red).} \label{Fig3}
\end{figure}

\emph{Band Structures and Berry Curvatures---.} We classify the twelve
atoms into three groups: (1) atoms (e.g. Sc, Mn, Fe) adsorbed at the hollow site, and graphene is magnetized; (2) atoms
(e.g. Ni) adsorbed at the hollow site, but graphene is not magnetized; (3) atoms (e.g. Cu) adsorbed at the top site.
To capture the physics arising from the Dirac points, we center
on the adsorption on the $4\times4$ supercells. The adsorption on
the $3\times3$ supercells will be discussed in the supplementary
materials.

Figure~\ref{Fig3} plots the band structures of adatom-graphene for
(a)~Sc, (b)~Mn, (c)~Fe, (d)~Cu, (e)~Ni adsorbed at the \emph{hollow}
sites and (f)~Cu adsorbed at the \emph{top} site. Panels ($\rm a_1$)-($\rm f_1$)
exhibit the full bulk bands along the high symmetry lines, and
panels ($\rm a_2$)-($\rm f_2$) zoom in the bands near the Dirac points $K$ and $K'$.
We see that sizable bulk band gaps open at the Dirac points in panels
($\rm a_2$)-($\rm d_2$) and ($\rm f_2$). The presence of such a bulk gap
indicates an insulating state. However, we note that all the Fermi levels lie outside
the bulk gaps except that in panel ($\rm c_2$) for Fe-adsorption. Thus, to realize this insulator one has to artificially adjust (e.g. by applying a gate voltage) the Fermi levels to be inside the gap.

To further identify the nontrivial topological properties of the obtained insulators, we turn to the Berry
phase analysis in the momentum space~\cite{NiuRMP}. By constructing the
Bloch wavefunctions $\psi$ from self-consistent potentials, the
Berry curvature $\Omega(\bm{k})$ can be obtained using the following
formula~\cite{Kubo1,Kubo2,Kubo3}:
\begin{eqnarray}
&&\Omega(\bm{k})=\sum_{n}f_n\Omega_{n}(\bm{k}) \nonumber\\
&&\Omega_n(\bm{k})=-{\sum_{n^{\prime} \neq n}} {\frac{2 {\rm {Im}}
\langle \psi_{n \bm{k}}|v_x|\psi_{n^\prime \bm{k}} \rangle \langle
\psi_{n^\prime \bm{k}}|v_y|\psi_{n \bm{k}} \rangle }
{(\omega_{n^\prime}-\omega_{n})^2}} \label{berry}
\end{eqnarray}
where the summation is over all $n$ bands below the bulk band gap,
$f_n$ is the Fermi-Dirac distribution function, $\omega_n\equiv E_n/
\hbar$, and $v_{x(y)}$ is the velocity operator. In panels ($\rm a_2$)-($\rm d_2$) and ($\rm f_2$), we also plot the Berry curvatures $\Omega(\bm{k})$ in
dotted curves. One can observe that in panels ($\rm a_2$)-($\rm d_2$) the Berry
curvatures $\Omega$ at $K$ point share the same sign as that at
$K^\prime$ point. By integrating the Berry curvatures over the first
Brillouin zone, it gives rise to a nonzero Chern number
$\mathcal{C}=+2$ for the insulating state of Sc, Fe, Cu
adsorption, and $\mathcal{C}=-2$ for the insulating state of the Mn
adsorption. The non-vanishing Chern number $\mathcal{C}$ in the absence of
magnetic field signifies a QAH state~\cite{Supp}, which was proposed
theoretically~\cite{haldane} but still not realized in experiment. In the subsequent section, we show the mechanism of the formation of the QAH effect in graphene.

Figure~\ref{Fig3} clearly shows that the adatoms of group (i) can open a non-trivial bulk band gap and achieve the QAH effect, but that of group (ii) can not. Therefore, we conclude that the magnetization in graphene, which breaks the time-reversal symmetry, is an essential ingredient for the QAH effect. Alternatively, we find that no bulk band gap opens for the group (i) atom-adsorption when
the spin-orbit coupling is switched off. This makes the spin-orbit coupling the other key ingredient of the QAH effect. From Table~\ref{table1}, one can notice that there is substantial charge transfer from the metal adatoms to graphene. Due to the asymmetry of the adatom-graphene structure, a sizable potential gradient is formed at the interface, which induces a considerable Rashba spin-orbit coupling~\cite{Rashba}. Panel ($\rm d_2$) for Cu adsorbed at the hollow site further supports that the 4$s$ spin-polarized electron transfer can also realize the QAH state. Therefore we summarize that the realization of the QAH state is only \emph{dependent} on the spin-polarized charge transfer, but \emph{independent} of the details in the adsorption (i.e. strong/weak hybridization). Although the induced Rashba spin-orbit coupling and magnetization in graphene are non-uniformly distributed in the structure, the gap opening mechanism near the Dirac points is similar to that of the tight-binding model with uniformly distributed Rashba spin-orbit coupling and magnetization~\cite{qiao}: first, the magnetization lifts the spin up and down bands, accompanying with the bands crossing; second, the Rashba spin-orbit coupling opens a bulk gap at the bands crossing points.

Because of the Van de Vaals force, the noble metal adatoms favor the top adsorption site. Besides the resulting Rashba spin-orbit coupling and exchange field, the major difference from the hollow-site adsorption is the extra introduction of the inequality of AB-sublattice potential, which breaks the spacial inversion symmetry. For example, we have considered the case with Cu adsorbed at the top site as shown in the panel ($\rm f_2$). Through plotting the Berry curvature distribution, we observe that $\Omega$ at $K$ point exhibits opposites sign as that at $K'$ point. This strongly signals a QVH state with canceling Chern numbers~\cite{Supp}, i.e. $\mathcal{C}=0$, but $\mathcal{C}_K=-\mathcal{C}_{k'}=1$. We can attribute this result to the leading role of the AB sublattice inequality suppressing the joint effect from Rashba spin-orbit coupling and exchange field.

\emph{Conclusions ---.} We study the magnetic and electronic
properties of graphene adsorbed with 12 metal atoms using first
principles calculation methods. We show that the QAH effect can be realized by doping the magnetic adatoms (i.e. Sc, Ti, V, Mn, Fe, Cu, Ag, and Au) at the hollow site of graphene, which originates from the joint effect between the Rashba spin-orbit coupling and magnetization. We also find that when the noble metal atoms (i.e. Cu, Ag, and Au) are adsorbed at the top site, the QVH effect can be observed due to the inversion symmetry breaking from the inequality of the AB sublattice potential. Our findings of the QAH and QVH states in the
adatom-graphene systems not only provides a new gap opening scheme
for the industrial application, but also pave the way for the
realization of the dissipationless charge/valley current in
spintronics and valleytronics.

\emph{Acknowledgement ---.} Z.Q. was supported by NSF (DMR 0906025)
and Welch Foundation (F-1255). Q.N. was supported by
DOE~(DE-FG03-02ER45958, Division of Materials Science and
Engineering) and Texas Advanced Research Program. Y.Y. was supported
by NSF of China (10974231) and the MOST Project of China
(2007CB925000, and 2011CBA00100). The Texas Advanced
Computing Center and Supercomputing Center of Chinese Academy of
Sciences are gratefully acknowledged for high performance computing
assistance.

$^\dag$ These authors contribute equally;

$^\ddag$ ygyao@aphy.iphy.ac.cn;

$^\S$ On leave from the University of Texas at Austin.

\begin{appendix} 
\section{Supporting Materials}

\title{Supplementary Material: Magnetic and Electronic Properties of Metal Atom Adsorbed Graphene}

In Section I, we shall discuss the band structures of the $3\times3$ supercells of graphene adsorbed with various transition metal atoms. Our results confirm that the quantum anomalous Hall (QAH) state can also be formed even though the non-trivial bulk gap opens at the $\Gamma$ point but not at the $K/K'$ points in the $3\times3$ supercells of graphene. In Section II, we introduce the definitions of the QAH effect and quantum valley Hall (QVH) effect.

\section{Band structures for adatom-$3\times3$ supercells of graphene}

In the main text, we have shown the band structures for the $4\times4$ supercells including the spin-orbit coupling and magnetization using the first-principles calculation method. The corresponding $K$ and $K'$ valleys are well-separated and distinguishable. We have shown that the resulting topological (QAH/QVH) states can be formed due to the non-trivial bulk gap opening around the $K$ and $K'$ points.

For the the adsorption in the $3\times3$ supercells of graphene, due to the folding of both $K$ and $K'$ into the same $\Gamma$ point, valley indices are indistinguishable. Therefore, it has not attracted so much interest as those structures (i.e. $4\times4$, $5\times5$, $7\times7$) with distinguishable valleys.
However, in experiment it may form the $3\times3$ supercell structures during the adsorption of diluted atoms on top of graphene. Therefore, it is necessary to know that whether the topological states ({i.e.} QAH state) proposed in the $4\times4$ supercells of graphene can still survive or not in the $3\times3$ supercells of graphene.

\begin{figure}
\includegraphics[width=8cm,totalheight=4cm,angle=0]{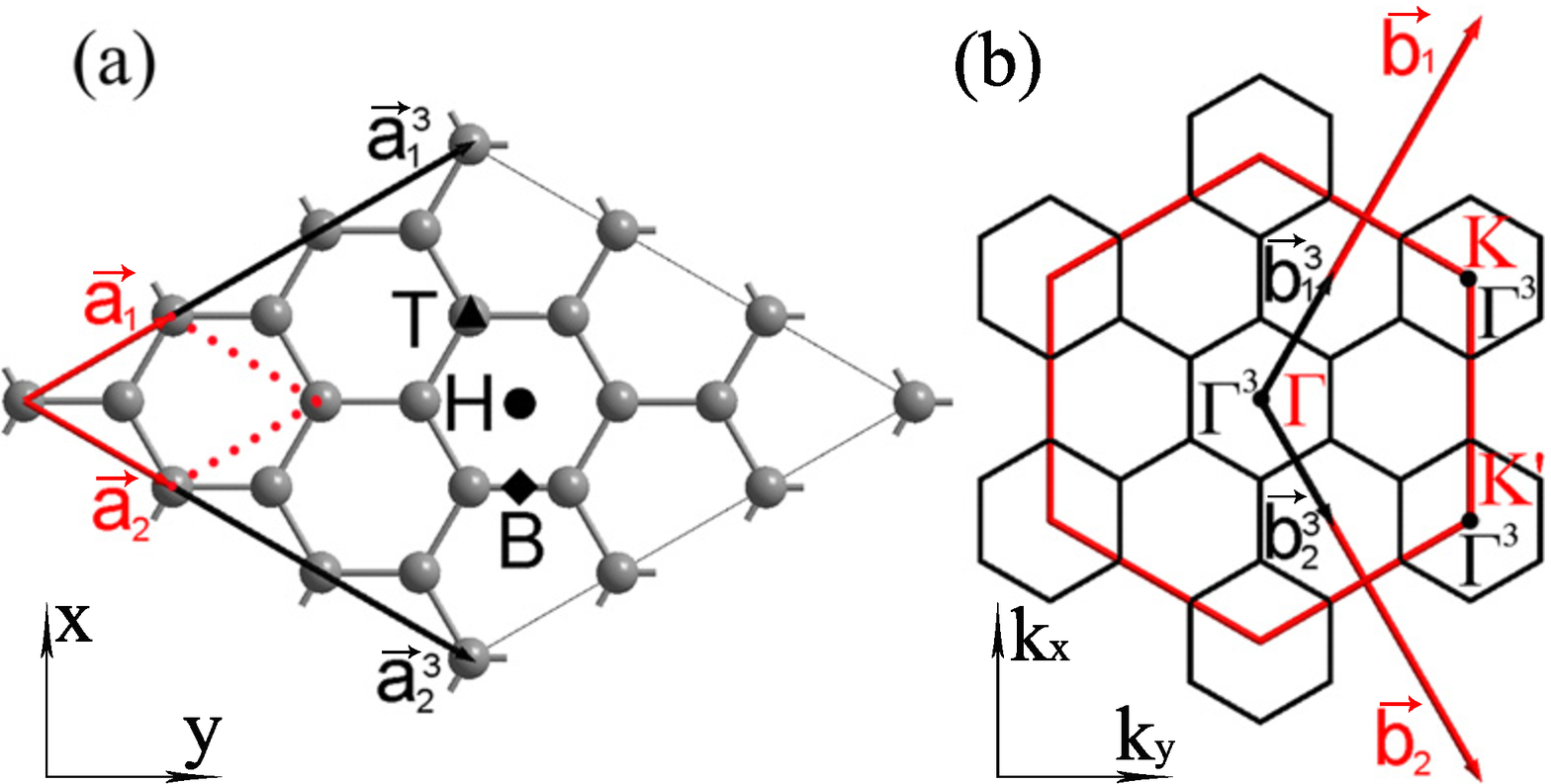}
\caption{(Color online) (a) A $3\times 3$ supercell of graphene.
$\vec{a}_{1,2}$ ($\vec{a}^3_{1,2}$) indicates the primitive vectors
for the $1\times 1$~~($3\times 3$) supercell. Three possible
adsorption sites of single adatom on graphene are labeled as:
hollow~(H), top~(T), and bridge~(B). (b) Corresponding reciprocal
momentum space structures: $\vec{b}_{1,2}$ and $\vec{b}^3_{1,2}$ are
for reciprocal vectors for the the $1\times 1$ and $3\times 3$
supercells, respectively. $K$ and $K'$ points for the $3\times3$
supercell of graphene are folded into the $\Gamma$ point.}
\label{sup1}
\end{figure}

In Fig.~\ref{sup1} (a) we schematically plot the $3\times3$
supercell of graphene compared to the $1\times1$ supercell of
graphene, with $\vec{a}_{1,2}$ and $\vec{a}^3_{1,2}$ representing the
primitive vectors for the $1\times1$ and $3\times3$ supercells.
Panel (b) illustrates the Brillouin zone of the $3\times3$ supercell
comparing to that of the $1\times1$ supercell. One can note that the original valleys $K$
and $K'$ are folded into the same $\Gamma$ point in the $3\times3$
supercell of graphene, which makes the two valleys indistinguishable.

\begin{figure*}
\includegraphics[width=14cm,totalheight=22cm,angle=0]{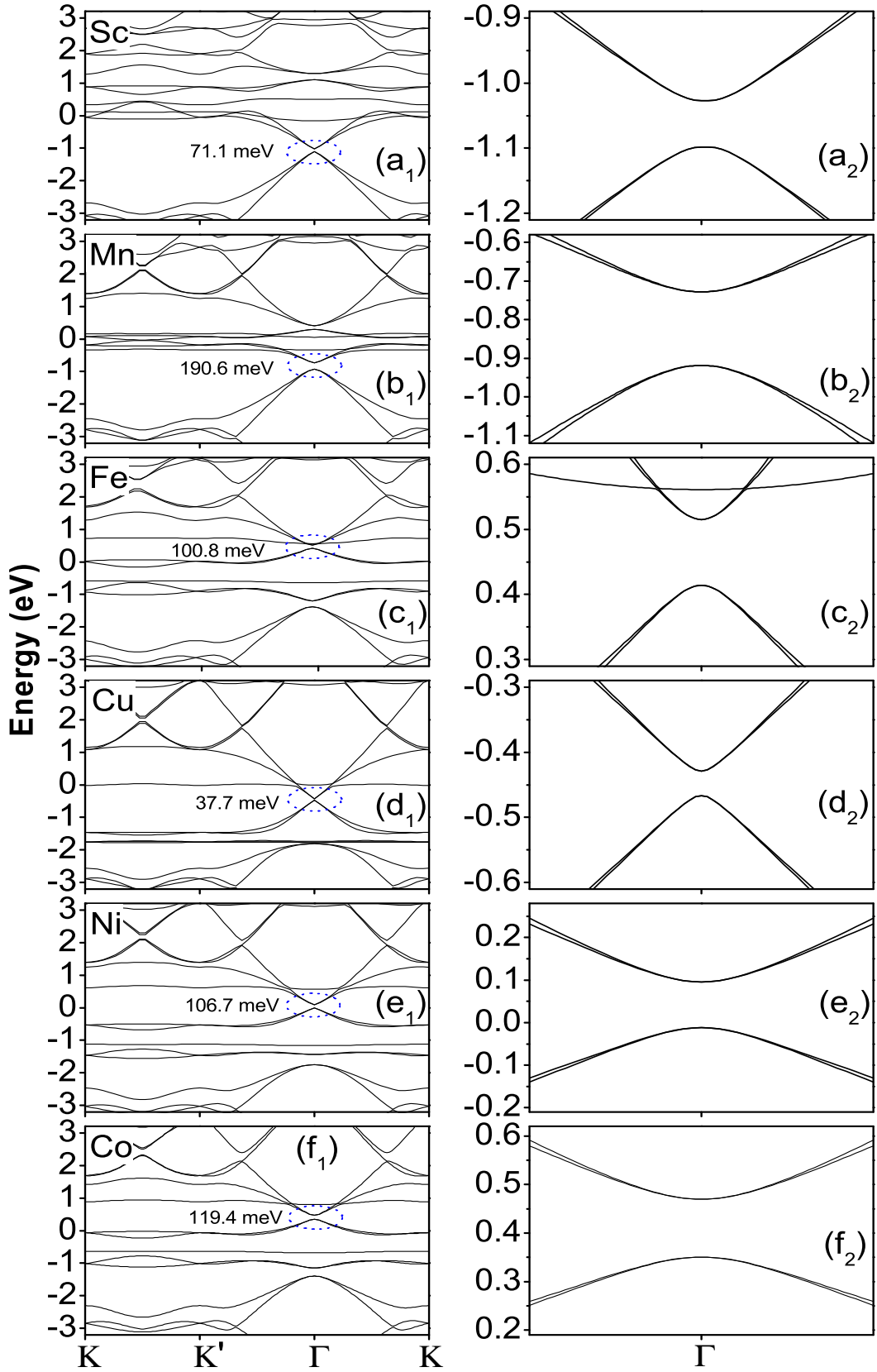}
\caption{(Color online) First column ($\rm a_1$-$\rm f_1$): full
band structures of adatom-$3\times3$ supercell of graphene for Sc,
Mn, Fe, Cu, Ni, Co adsorbed on top of the hollow position without
including magnetization and spin-orbit coupling along the high
symmetry line. Second column ($\rm a_2$-$\rm f_2$): Zooming in of
the circled bands around $\Gamma$ point in panels ($\rm a_1$-$\rm
f_1$). Bulk band gaps are opened at the $\Gamma$ point due to the
inter-valley scattering effect. Note that there are two equivalent
band gaps at $\Gamma$ point except in the band structure of
Cu-adsorption.} \label{sup2}
\end{figure*}

\begin{figure*}
\includegraphics[width=14cm,totalheight=10cm,angle=0]{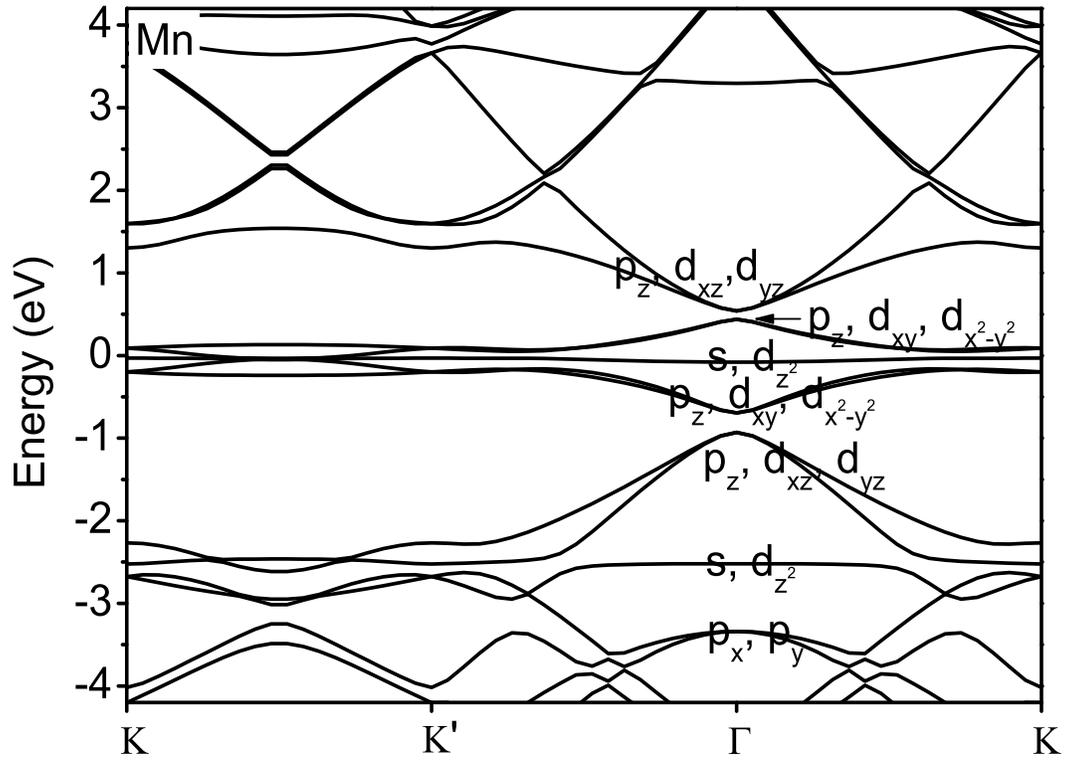}
\caption{(Color online) Projected bands analysis for the bands near
the two bulk band gaps in Figure \ref{sup2}($\rm b_1$). We can see
that the bands near the two bulk gaps possess the same orbital
components.} \label{sup3}
\end{figure*}

Figure~\ref{sup2} exhibits the band structures of $3\times3$
supercells of graphene adsorbed with (a) Sc, (b) Mn, (c) Fe, (d) Cu,
(e) Ni and (f) Co adatoms on top of the hollow position without
involving spin degrees of freedom, i.e. nonmagnetic calculation. In the first column~[\emph{i.e.}
panels ($\rm a_1$)-($\rm f_1$)], we plot the full band structures
along the high symmetry lines. From the band structures shown in the
second column~[{i.e.} panels ($\rm a_2$)-($\rm f_2$)], the
zooming in of the circled bands in the first column, we observe that
large band gaps can open at the $\Gamma$ point due to the
inter-valley scattering between $K$ and $K'$ valleys, {i.e.}
$\Delta=71.1,190.6,~100.8,~37.7,~106.7,~119.4$ in units of $meV$ for
the $3\times3$ supercell of graphene adsorbed with Sc, Mn, Fe, Cu,
Ni, and Co, respectively. This is completely distinct from that of
the adsorption on the $4\times4$ supercells of graphene in the main
text, where the perfect Dirac-cone dispersion holds at both $K$ and
$K'$ points. In our calculations we have
used the Hydrogen atom to saturate the dangling bond of Mn atom, because in the Mn-adsorption the dangling d-band is properly
located inside the resulting bulk gap. In the first column one can see that there are two bulk band gaps opened at
the $\Gamma$ points. As shown in Fig.~\ref{sup3} the projected band
analysis points out that the two gaps are essentially equivalent by
combining the same orbital components of $P_z$, $d_{xz}$ and
$d_{yz}$ (or $P_z$, $d_{xy}$ and $d_{x^2-y^2}$).

\begin{figure*}
\includegraphics[width=14cm,totalheight=22cm,angle=0]{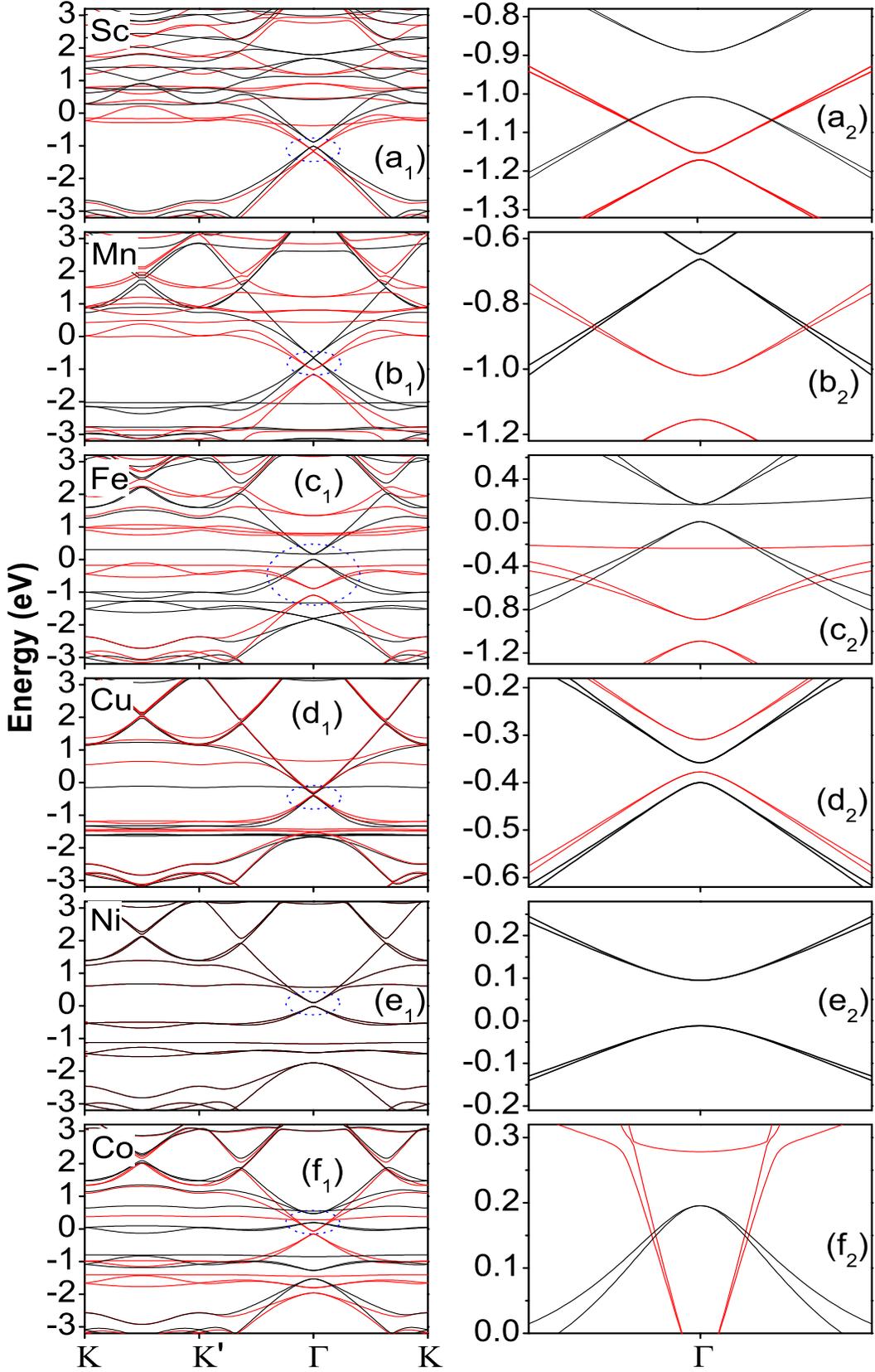}
\caption{(Color online) First column ($\rm a_1$-$\rm f_1$): full
band structures of adatom-$3\times3$ supercell of graphene for Sc,
Mn, Fe, Cu, Ni, Co adsorbed on top of the hollow position by
including the magnetization along the high symmetry line. Second
column ($\rm a_2$-$\rm f_2$): Zooming in of the circled bands around
$\Gamma$ point in panels ($\rm a_1$-$\rm f_1$). Colors are used to
distinguish the spin-up (black) and spin-down (red) bands. In the
second column all the spin-up and spin-down bands are crossing due
to the large magnetization except that for Cu and Ni adsorption. In
panel ($\rm d_2$) for Cu-adsorption the magnetization is less than
the original band gap amplitude, therefore the relative shift
between the spin-up and spin-down bands does not cross to close the
band gap. In panel ($\rm e_2$) for Ni-adsorption the band structure
is exactly the same as that in Figure \ref{sup2} (e) because of the
vanishing magnetic moment of Ni-adatom.} \label{sup4}
\end{figure*}

Similar to the adsorption on the $4\times4$ supercell of graphene, when the
magnetization is included, the spin up (down) bands are upward
(downward) shifted (see Figure \ref{sup4}). This results in three
possible results: (i) As plotted in panels (a$_2$)-(c$_2$) and (f$_2$)
for Sc, Mn, Fe, and Co adsorption, the induced magnetization is so
large ({i.e.} $\lambda$=0.213, 0.432, 1.080, 0.442 in units of eV) that the original band gaps from the inter-valley scattering
are closed due to the crossover between the spin-down bands from the
conduction bands and the spin-up bands from the valence bands. (ii)
For panel (d$_2$) of Cu-adsorption, the magnetization is moderately
large ({i.e.} $\lambda$=0.035 eV) that the band gaps are just slightly decreased but not closed.
(iii) For the Ni-adsorption in panel (e$_2$) the band structure is
exactly the same as that shown in Fig.~\ref{sup2} (e$_2$) because
there is no magnetization of graphene induced from the Ni
adsorption.

\begin{figure*}
\includegraphics[width=14cm,totalheight=22cm,angle=0]{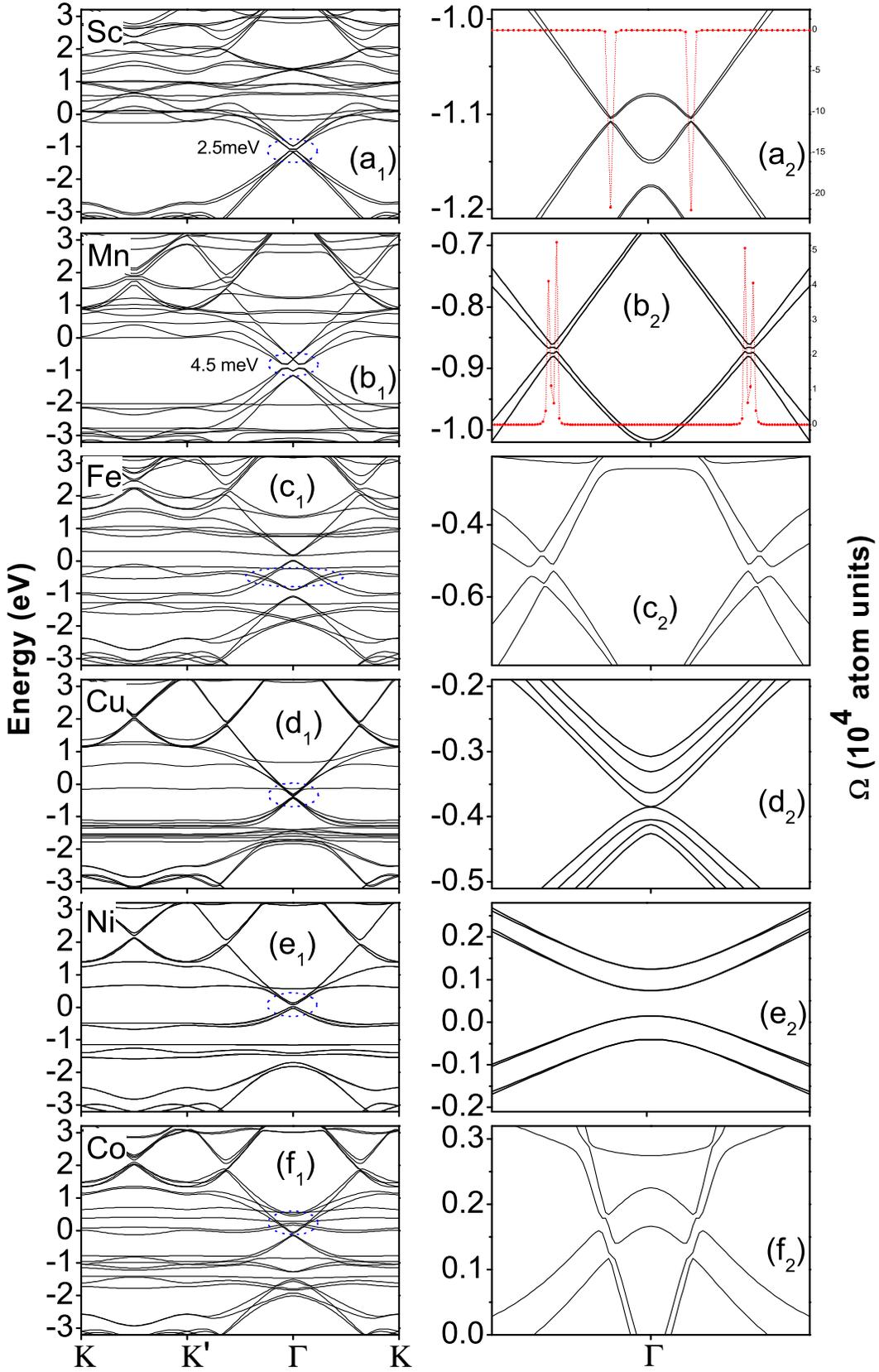}
\caption{(Color online) First column ($\rm a_1$-$\rm f_1$): full
band structures of adatom-$3\times3$ supercell of graphene for Sc,
Mn, Fe, Cu, Ni, Co adsorbed on top of the hollow position by
including both the magnetization and spin-orbit coupling along the
high symmetry line. Second column ($\rm a_2$-$\rm f_2$): Zooming in
of the circled bands around $\Gamma$ point in panels ($\rm a_1$-$\rm
f_1$). In panels ($\rm a_2$-$\rm c_2$) and ($\rm f_2$), nontrivial
bulk band gap open at the crossing points. However, only bulk band
gaps are globally opened in panels ($\rm a_2$-$\rm b_2$) for Sc, Mn
adsorption, while the gaps in panels ($\rm c_2$) and ($\rm f_2$) are
just locally opened. Therefore only the Sc/Mn adsorbed $3\times3$
graphene structure can exhibit the nontrivial insulating properties
when the Fermi-level is artificially adjusted to the bulk band gaps.
In addition, we also plot the total Berry curvatures $\Omega(k)$ in
panels ($\rm a_2$-$\rm b_2$) for all the valence bands below the
bulk band gaps.} \label{sup5}
\end{figure*}

When the spin-orbit coupling is further considered, we find that
non-trivial band gaps can open at the crossing points between the
spin-up and spin-down bands [see panels ($\rm a_2$)-($\rm c_2$) and
($\rm f_2$) in Fig.~\ref{sup5}], which shows similar origin as that
in the adsorption on the $4\times4$ supercells in the main text.
However, we find that only the bulk band gaps in
panels ($\rm a_1$-$\rm b_1$) are globally opened for Sc ($\Delta=2.5$ meV) and Mn ($\Delta=4.5$ meV), while that for Fe
and Co in panels ($\rm c_1$) and ($\rm f_1$) are just locally opened.
Therefore in the following, we will only focus on the nontrivial
topological properties for Sc and Mn adsorption. Through calculating
the Berry curvatures $\Omega$ along the high symmetry lines for all
the valence bands below the bulk gap, we find that all the Berry
curvatures $\Omega$ share the same sign as shown in panels ($\rm a_2$) and ($\rm b_2$), resulting in the
non-vanishing Chern number $\mathcal{C}$. By integrating all the
Berry curvatures below the band gaps, our calculation shows that
$\mathcal{C}=+2/-2$ for the Sc/Mn adsorption. This again
signals a QAH insulating state for the
energies inside the bulk gap which will be explained in detail
in Section II. Since the Fermi energy is not located inside the band gap, one has to
apply external gate voltage to adjust the Fermi level inside the
band gap to realize the insulating state. While for the other two
adsorption of Cu and Ni shown in panels (d2)-(e2), we can see that
the spin-orbit coupling further decreases the bulk gaps.

In summary, using the first-principles calculation methods, we find
that for the $3\times3$ supercell of graphene adsorbed with atoms
(i.e. Sc, Mn, Fe, Cu, Ni) on the top of hollow position, trivial
bulk band gaps can open at the $\Gamma$ point due to the
inter-valley scattering without considering the spin degrees of
freedom. For some particular adatoms ({i.e.} Sc, Mn) when the
magnetization and spin-orbit coupling are involved, we find that
non-trivial bulk band gaps can open around the $\Gamma$ point to
realize the QAH effect. This finding not only generalizes our proposal of QAH effect in the $4\times4$ supercell to the $3\times3$ supercell of graphene, but also enhances the experimental realizability of the first QAH effect in graphene.

\section{Definitions of Quantum Anomalous Hall and Quantum valley Hall effects}

\subsection{Quantum Anomalous Hall Effect}
The breaking of the time-reversal symmetry is a necessary condition
for the formation of the Hall effects. In general, there are two
kinds of methods breaking the time-reversal symmetry: the external
magnetic field and the intrinsic magnetization of the ferromagnetic
materials. When the two-dimensional electron gas is subjected to a
strong perpendicular magnetic field, Laudau-levels can be formed to
exhibit the integer quantum Hall effect with the Hall conductance
exactly quantized in units of $e^2/h$~\cite{Klitzing}. The precise
quantization of Hall conductance can be understood in terms of a
topological invariant known as the Chern number $\mathcal{C}$, which
must be an integer~\cite{PhysicsToday}. Therefore the integer
quantum Hall conductance can be expressed as:
\begin{equation}
\sigma_{xy}= \mathcal{C} \frac{e^2}{h}
\end{equation}

Contrary to the ordinary Hall effect due to the magnetic field, the
intrinsic magnetization induced Hall effect is termed as anomalous
Hall effect. Though this anomalous Hall phenomenon has been known
for over one century, its origin is still unclear~\cite{AHE1,AHE2}.
The quantized version of the anomalous Hall effect, also named as
quantum anomalous Hall (QAH) effect, was first theoretically
proposed by Haldane in a Honeycomb toy model after the experimental
observation of the integer quantum Hall effect in
1980s~\cite{haldane}. However, the QAH effect is still not observed
in experiment.

In our work on the metal adatom-graphene structures, we find that
the QAH state can be observed in both $3\times3$ and $4\times4$
supercells of graphene. In the main text the bulk gaps are opened at
both $K$ and $K'$ points when the magnetization and spin-orbit
coupling are considered. Because the Berry curvatures near $K$ and
$K'$ for the $4\times4$ supercell show the same sign, therefore the
summation of the Berry curvatures of the whole Brilloin-zone is finite, which indicates that the Chern number must be a
nonzero integer. Since we do not apply any external magnetic field,
therefore we can conclude this non-vanishing Chern number
corresponds to the QAH state. Surprisingly, in the Section I of this
supplementary materials we further find that the adsorption on the
$3\times3$ supercell can also contribute to the QAH effect through
opening a nontrivial bulk band gap near the $\Gamma$ point by
involving the magnetization and spin-orbit coupling.

\subsection{Quantum Valley-Hall Effect}
Quantum valley Hall (QVH) effect is named after the integer
quantized but canceling quantum-Hall responses at $K$ and $K'$
valley points. This effect can be arisen from the inversion-symmetry
breaking, \emph{i.e.} applying staggered AB sublattice potentials on
the single layer graphene~\cite{QVH3}, or considering gated bilayer
graphene~\cite{QVH1}. In terms of Chern number representation,
$\mathcal{C}_K=-\mathcal{C}_{K'} \neq 0$ but the total Chern number
of the system is $\mathcal{C}=\mathcal{C}_K+\mathcal{C}_{K'}=0$. One
has to note that there is no edge modes for the QVH effect in the
single layer graphene~\cite{QVH2,QVH3}, while for the bilayer
graphene there is indeed gapless edge states inside the bulk band
gap~\cite{QVH4}. These edge states are robust against smooth-type
disorders, which is protected by the large momentum separation
between two valleys.

In our studied adatom-$3\times3$/$4\times4$ structures, no QVH state
can exist in the $3\times3$ supercell systems because of the
inter-valley scattering which makes valleys indistinguishable.
In the $4\times4$ supercell with noble atom adsorbed graphene
systems, our calculation shows that the Berry curvatures at $K$ and
$K'$ are non-vanishing and opposite, which indicates that valleys
$K$ and $K'$ correspond to two opposite nonzero Chern numbers. Based
on the definition of QVH effect, we can claim that our finding of
the noble adatom adsorption on top of the $4\times4$ supercell leads
to the QVH effect.

\end{appendix}

\begin{thebibliography}{99}
\bibitem{exchangefield}
It is obtained from $\lambda=(\varepsilon_\uparrow - \varepsilon_\downarrow)/2$ at $K/K'$ point.
\end{thebibliography}
\end{document}